\documentclass[%
jcp,
 aip,
jmp,%
 sd,%
 amsmath,amssymb,
preprint,%
]{revtex4-1}

\usepackage{xcolor}
\usepackage{graphicx}
\usepackage{dcolumn}
\usepackage{bm}
\usepackage{hyperref}
\usepackage{multirow}
\usepackage{etoolbox}

\begin{document}


\title[]{Perspective: Nonequilibrium dynamics of localized and delocalized excitons in colloidal quantum dot solids}

\author{Elizabeth M. Y. Lee}
\affiliation{%
 Department of Chemical Engineering, Massachusetts Institute of Technology,
\\Cambridge, Massachusetts 02139, USA
}%
\author{William A. Tisdale}
\email{tisdale@mit.edu}
\affiliation{%
 Department of Chemical Engineering, Massachusetts Institute of Technology,
\\Cambridge, Massachusetts 02139, USA
}%

\author{Adam P. Willard}
\email{awillard@mit.edu}
\affiliation{%
 Department of Chemistry, Massachusetts Institute of Technology,
\\Cambridge, Massachusetts 02139, USA
}%


\begin{abstract}
Self-assembled quantum dot (QD) solids are a highly tunable class of materials with a wide range of applications in solid-state electronics and optoelectronic devices.  In this perspective, we highlight how the presence of microscopic disorder in these materials can influence their macroscopic optoelectronic properties. Specifically, we consider the dynamics of excitons in energetically disordered QD solids using a theoretical model framework for both localized and delocalized excitonic regimes. In both cases, we emphasize the tendency of energetic disorder to promote nonequilibrium relaxation dynamics and discuss how the signatures of these nonequilibrium effects manifest in time-dependent spectral measurements. Moreover, we describe the connection between the microscopic dynamics of excitons within the material and the measurement of material specific parameters, such as emission linewidth broadening and energetic dissipation rate. 
\end{abstract}

\maketitle
\section{Introduction}
The optoelectronic properties of colloidal quantum dots (QDs) depend sensitively on their size, shape, and chemical composition.~\cite{Murray1995,Alivisatos1996}  This dependence has inspired the development of a class of solid materials made up of self-assembled QDs that exhibit highly tunable optoelectronic properties. 
This tunability has been leveraged to enable a wide range of solid-state applications, such as light-emitting diodes (LEDs),~\cite{Shirasaki2013,Dai2014} solar cells,~\cite{Sanehira2017} lasers,~\cite{Klimov2000} photodetectors,~\cite{McDonald2005,DeIacovo2016} and luminescent solar concentrators.~\cite{Coropceanu2014} Notably, the optoelectronic properties of QD solids can also depend on the spatial arrangement of QDs within the material. However, understanding this dependence has been a challenge because it conveys through collective interactions that are especially sensitive to heterogeneity in the QD population, arising due to the process of QD synthesis, and in the spatial distribution of QDs, arising due to the process of self-assembly.

In a typical QD solid, excited electrons and holes tend to localize on individual QDs. These oppositely charged carriers can co-localize on each QD to form neutral quasiparticles known as excitons. The dynamics of a localized exciton in a QD solid involves a series of hops, whereby the exciton moves from one QD to another through resonant energy transfer process. The theory to describe this type of exciton dynamics is that of F\"orster resonance energy transfer or FRET.~\cite{Forster1948,Forster2012} Many studies, both experimental and theoretical, have used FRET to gain a better understanding of the role of exciton dynamics in the macroscopic properties of QD solids.~\cite{Kagan1996,Crooker2002,Achermann2003,Kim2008,Lingley2011,Miyazaki2012,Poulikakos2014,Mork2014,Zheng2014,Wang2017} 

The limitations in energy transport that are implied by FRET-like exciton dynamics can be overcome by exploiting the quantum mechanical effect of exciton delocalization. When an exciton delocalizes over many individual QDs, it can more readily explore space and can undergo enhanced supertransfer through the constructive interference of individual QD transition dipole moments. 
Unfortunately, achieving exciton delocalization in QD solids has emerged as a significant challenge due to the persistence of weak inter-QD electronic coupling. This weak coupling is a consequence of insulating ligands that passivate QD surfaces as well as mismatched energetic resonances arising through variations in QD sizes. 
Theoretical simulations and model studies are essential to developing QD solids that support delocalized excitations.  

Here we present a general model framework, based on the principles of F\"orster theory, for simulating exciton dynamics in QD solids. We use this framework to study the effects of energetic disorder on the dynamics of localized and delocalized excitons. This model demonstrates how energetic disorder leads to nonequilibrium effects, and how those effects manifest in experiment. 
We then apply this framework to analyze spectrally-resolved transient photoluminescence measurements of CdSe QD solids. We then identify how material properties of QDs, such as energetic disorder, energy dissipation upon optical excitation, and emission linewidth broadening, can influence exciton diffusivity as well as the transient shift in average emission energy.
By extending this framework to the case where inter-QD coupling is larger, we are able to highlight the enhancement in exciton transport that can arise through exciton delocalization. Finally,  we summarize and discuss the potential of delocalization to improve the performance of technologies based on QD-solids. 


\section{Model of energy transfer in quantum dot solids}

\begin{figure}
\includegraphics{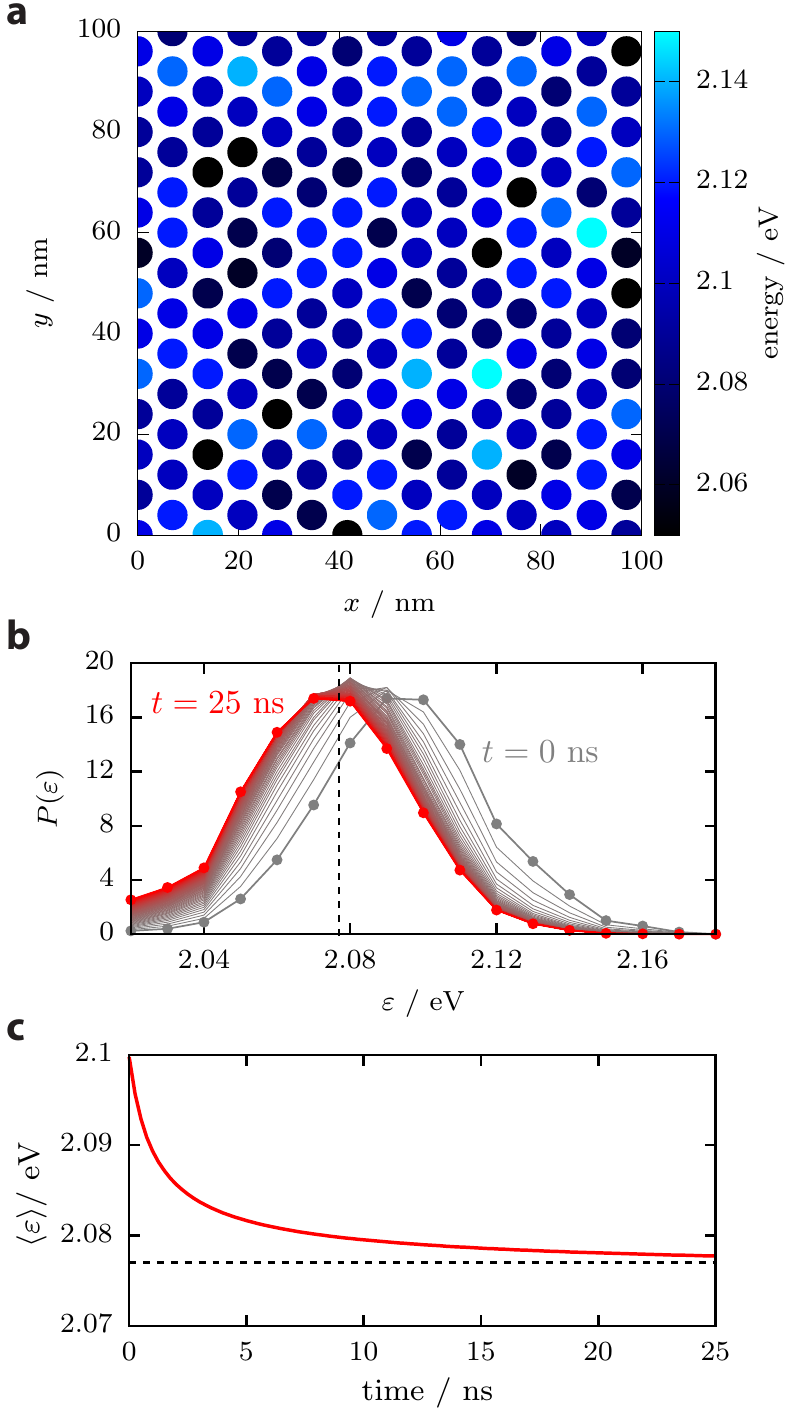}
\caption{Simulated exciton dynamics within energetically disordered quantum dot solid. (a) A configuration of a two-dimensional model of QD solid with nearest-neighbor distance of 8 nm. Point shading indicates the excitation energy of each model QD (b) Time evolution of the probability density distribution of exciton energy $\varepsilon$. Time points are indicated by its corresponding color. (c) Red line indicates the ensemble-averaged excitation energy, and the dashed black line correspons to the Boltzmann-weighted average over the ensemble of QD energies. 
Exciton dynamics is simulated via the chemical master equation with a set of parameters corresponding to CdSe quantum dots capped with organic ligands~\cite{Akselrod2014} ($\bar \varepsilon$ = 2.1 eV, $\sigma_{\mathrm{ih}}$ = 20 meV, $\sigma_\mathrm{h}$ = 30 meV, $\Delta_{\mathrm{ss}}$ = 40 meV, $n$ = 1.7, $\eta$ = 1,  $\tau$ = 10 ns, and $N=2500$) }
\label{fig:fig_ME}
\end{figure}

Exciton dynamics in QD solids can be understood by considering how an exciton is transferred from one QD to another. F\"orster theory provides a basis for describing this process in the incoherent limit.~\cite{Forster1948,Forster2012} In F\"orster theory, the coupling that drives energy transfer originates from the interaction of the transition dipole vectors of the donor and acceptor QDs. Thus, the rate of energy transfer between any two QDs scales as $1/d^6$, where $d$ is the seperation distance between the QDs.~\cite{Kim2008,Lingley2011} Due to the detailed balance condition, F\"orster theory also predicts faster rates for exciton transfer that is downhill in energy, leading to a transient redshift of the average emission energy in inhomogeneously broadened QD solids.~\cite{Akselrod2014,Lee2015a} Such variations in QD energies has been attributed to size, shape, and stoichiometric variations between individual QDs.~\cite{Cui2013,Utzat2017} This transient redshift has been observed in QD solids via spectrally-resolved transient photoluminescence measurements.~\cite{Kagan1996,Crooker2002,Achermann2003,Miyazaki2012,Mork2014,Poulikakos2014} 

The downhill energetic migration of excitons in QD solids has more subtle dynamic consequences due to the fact that thermalized excitons have, on average, fewer possible downhill transitions than non-thermalized excitons. This can lead to an exciton diffusivity that decreases over time, which has been observed in studies of time-resolved optical microscopy applied to inhomogeneously broadened QD solids,~\cite{Akselrod2014} agreeing with theoretical predictions of incoherent transport over disordered energy landscape.~\cite{Lee2015a} These experiments reveal that F\"orster theory correctly predicts scaling parameters that affect the energy transfer rate in QD solids.~\cite{Kim2008,Lingley2011,Mork2014,Akselrod2014,Kholmicheva2017} 

To understand the dynamics of excitons in QD solids, let us consider a model QD solid system that is illustrated in Figure~\ref{fig:fig_ME}a. This model system includes QDs assembled in a two-dimensional hexagonally closed packed lattice. QDs are inhomogeneously broadened such that exciton energy at a given QD site is drawn randomly from a Gaussian distribution with mean $\bar \varepsilon$ and standard deviation $\sigma_\mathrm{ih}$ (the inhomogeneous linewidth). Moreover, each QD is assigned a fixed transition dipole vector, $\hat \mu$, oriented randomly on the surface of a unit sphere, assuming that orientations of the transition dipoles moments are isotropic. 

According to FRET, the transition probability per unit time is given by,
\begin{equation}
k_{\mathrm{D} \to \mathrm{A}}=\frac{1}{\tau}\left(\frac{{R_0}}{{d_{\mathrm{DA}}}}\right)^6,
\label{eq:1}
\end{equation}
where $d_{\mathrm{DA}}$ is the center-to-center distance between donor (D) and acceptor (A) quantum dots, $\tau$ is the total lifetime of the donor quantum dot, and $R_0$ is the F\"orster radius. The dependence of exciton hopping rate on the excitation energies of QDs is contained within the the F\"orster radius expression,
\begin{equation}
{R_0}^6 = \frac{9}{{8\pi }}\frac{{{c^4}{\hbar ^4}}}{{{n^4}}}\eta {\kappa ^2}\int {\frac{{{\sigma }\left( \varepsilon ; \varepsilon_A \right){f}\left( \varepsilon ; \varepsilon_D \right)}}{{{\varepsilon ^4}}}} d\varepsilon,
\label{eq:radius}
\end{equation}
where $n$ is the refractive index; $c$ is the speed of light; $\hbar$ is the reduced Planck constant;  $\eta$ is the photoluminescence quantum yield; and $\kappa$ is the transition dipole orientation factor, calculated as $\kappa = \hat \mu_{\mathrm{D}} \cdot \hat \mu_{\mathrm{A}}-3(\hat \mu_{\mathrm{D}} \cdot  \hat d_{\mathrm{DA}})(\hat \mu_{\mathrm{A}} \cdot  d_{\mathrm{DA}})$, where $ \hat d_{\mathrm{DA}}$ is a unit vector pointing from the donor to the acceptor QDs. The integral term in Eq.~\ref{eq:radius} is the spectral overlap between the normalized emission spectrum of the donor, $f(\varepsilon;\varepsilon_\mathrm{D})$,  and the absorption spectrum of the acceptor, $\sigma(\varepsilon;\varepsilon_\mathrm{A})$. 

Assuming that each absorption and emission lineshape is a Gaussian with a standard deviation of $\sigma_\mathrm{h}$ and that the $\varepsilon^{4}$ term in Eq.~\ref{eq:radius} varies slowly over integral,~\cite{Ahn2007} we can simplify Eq.~\ref{eq:radius} as,
 \begin{eqnarray}
R_0^6=  \frac{C \kappa^2}{\left[\frac{1}{2} \left( \varepsilon_\mathrm{D} + \varepsilon_\mathrm{A} - \Delta_{\mathrm{ss}}\right) \right]^4}\exp{\left[-\frac{ \left(\varepsilon_\mathrm{D} - \varepsilon_\mathrm{A} -\Delta_\mathrm{ss}\right)^2 }{4{\sigma_\mathrm{h}}^2}\right]},
\label{eq:radius_new}
\end{eqnarray}
where $C$ is a collection of physical constants,
\begin{equation}
C\equiv \frac{9}{8 \pi}\frac{c^4 \hbar^4}{n^4} \eta \sigma,
\end{equation}
and $\Delta_\mathrm{ss}$ is Stokes shift, which is the difference in energy between the absorption and emission energy peaks due to the rapid electronic and nuclear relaxation that follows the excitation. 

By applying the FRET rate equation (Eq.~\ref{eq:1}) to the model system depicted in Figure~\ref{fig:fig_ME}a, it is possible to generate an entire transition rate matrix for excitons within the model QD solid. This rate matrix can then be used to simulate the energy transport properties of model materials or to aid in the interpretation of experiments. To accomplish this, we use chemical master equation as described in the following section.

\section{Localized exciton hopping picture}
Incoherent energy transport process in which localized excitation jumps from site to site can be modeled using the chemical master equation,
\begin{equation}
\frac{d P_i}{dt} = \sum_{j \neq i}k_{j \to i} P_j - \sum_{i} k_{i \to j} P_i ,
\label{eq:me}
\end{equation}
where $P_i$ is the exciton probability density at site $i$. Figure~\ref{fig:fig_ME}b shows the numerically exact solution of Eq.~\ref{eq:me}, in which QDs are first uniformly excited, \textit{i.e.}, $P_i = 1/N, \forall i$, and non-interacting excitons undergo an energetic relaxation within a Gaussian density of states. The parameters used in this calculation were based on those estimated for colloidal CdSe QDs that were experimentally studied by Akselrod, \textit{et al.}~\cite{Akselrod2014}

As seen in Figure~\ref{fig:fig_ME}b, the exciton dissipates its energy over time as indicated by the redshift in the average energy, $\langle \varepsilon \rangle$, while the probability distribution maintains the initial Gaussian shape with a width of $\sigma_\mathrm{ih}$. Additionally, Figure~\ref{fig:fig_ME}c shows that the mean energy saturates after long time, indicating that excitons reach a dynamic equilibrium if their lifetimes were infinitely long. Based on the approximations we made to derive a closed form expression for the energy transfer rate (Eq.~\ref{eq:radius_new}),~\cite{Ahn2007} we present an analytical expression for average energy at equilibrium in the case of FRET for the first time. We recall that at equilibrium, exciton population satisfies the detailed balance condition,
\begin{equation}
w(\varepsilon_i) k_{i \to j} = w(\varepsilon_j) k_{j \to i},
\end{equation}
where $w(\varepsilon)$ is the exciton probability density at equilibrium. 
The equilibrium probability density obeys the relation,
\begin{equation}
\frac{w(\varepsilon_j)}{w(\varepsilon_i)} = \frac{k_{i \to j}}{k_{j \to i}} =  \exp{\left[-\frac{\Delta_\mathrm{ss} \left(\varepsilon_j - \varepsilon_i \right) }{{\sigma_\mathrm{h}}^2}\right]},
\end{equation}
such that, 
\begin{equation}
w(\varepsilon) \propto \exp{\left(-{ \Delta_\mathrm{ss} \varepsilon }/{{\sigma_\mathrm{h}}^2}\right)},
\end{equation}
which takes the analogous form of a canonical distribution if $\sigma_\mathrm{h}^2/\Delta_\mathrm{ss} = k_\mathrm{B} T$, where $k_\mathrm{B}$ is the Boltzmann constant and $T$ is temperature. 

If the initial probability density distribution is $P_0(\varepsilon)$, then the final probability density distribution of exciton is given by $P_\infty(\varepsilon) \propto P_0(\varepsilon)w(\varepsilon)$. Therefore, if the initial exciton distribution obeys Gaussian statistics with a mean $\bar \varepsilon$ and a standard deviation $\sigma_\mathrm{ih}$, then $P_\infty$ will also be a Gaussian,
\begin{equation}
P_\infty(\varepsilon)\propto \exp{\left[-\frac{( \varepsilon-\bar \varepsilon + \Delta_\mathrm{ss} \left(\sigma_\mathrm{ih}/\sigma_\mathrm{h})^2 \right)^2}{2{\sigma_\mathrm{ih}}^2 } \right]},
\end{equation}
with the same linewidth as before but centered at $\varepsilon = \bar \varepsilon - \Delta_\mathrm{ss}\left({\sigma_\mathrm{ih}}/{\sigma_\mathrm{h}}\right)^2$. The loss in excitation energy upon exciton migration is given by,
\begin{equation}
\bar \varepsilon - \langle \varepsilon_\infty \rangle = \Delta_\mathrm{ss}\left(\frac{\sigma_\mathrm{ih}}{\sigma_\mathrm{h}}\right)^2. 
\label{eq:eqm}
\end{equation}
Thus exciton population at equilibrium reaches a final average energy that is determined by the site energy disorder ($\sigma_\mathrm{ih}$), the available thermal energy ($\sigma_\mathrm{h}$), and the reorganization energy ($\Delta_\mathrm{ss}$). Figure~\ref{fig:fig_ME}c reveals that even with a finite exciton lifetime ($\tau = 10$ ns), the average energy relaxes to the value predicted by Eq.~\ref{eq:eqm} for parameters used to model CdSe QDs capped with organic ligands.~\cite{Akselrod2014}

\begin{figure}
\includegraphics{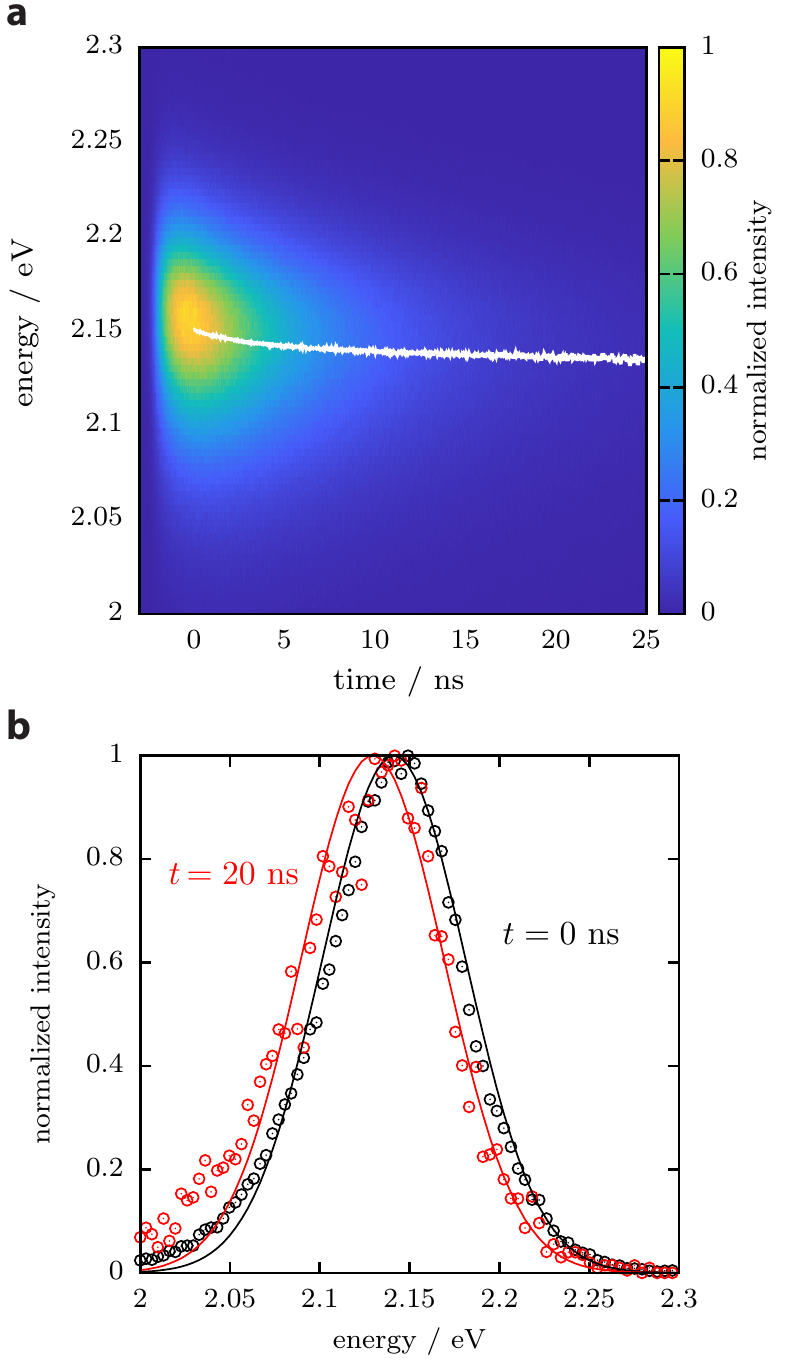}
\caption{Spectrally resolved transient photoluminescence measurements of CdSe quantum dot solid. (a) Energetic relaxation of non-interacting population of exciton within an inhomogeneously broadened ensemble. Solid white line indicates peak emission energy as a function of time. Emission intensity has been normalized such that the maximum intensity at $t=0$ equals 1. (b) Emission spectra at initial and at steady state ($t \approx 20$ ns) that were normalized with respect to the maximum intensity measured at each time point. Circles are experimental data points, and the lines are fits to the data with a Gaussian function.}
\label{fig:fig_exp}
\end{figure}

We apply our model to analyze spectrally resolved transient photoluminescence measurements of CdSe/ZnCdS core-shell colloidal QD assembly, as shown in Figure~\ref{fig:fig_exp}. For details about the sample and the measurement technique, we refer readers to \citet{Akselrod2014}. Briefly, QDs are excited at 405 nm (3.06 eV) diode laser producing pulses $\sim 500$ ps in duration with a repetition rate of 10 MHz and a low laser fluence to probe dynamics of non-interacting excitons. As seen in Figure~\ref{fig:fig_exp}b, emission spectra of this QD sample have an asymmetric lineshape, with an elongated tail toward low energy ($< 2.06$ eV). Moreover, the ratio of the photoluminescence intensity at the low energy tail to that at high energy (between 2.1 and 2.2 eV) increases over time. This asymmetric emission lineshape has been also observed in other colloidal QD systems, in which the low energy tail is attributed to sub band-edge states whose origin is under debate.~\cite{Caram2016}

In our analysis, we only consider the band-edge exciton state whose emission peak is fit to a normal distribution as illustrated in Figure~\ref{fig:fig_exp}. This emission peak has a total linewidth of about 28 meV that stays relatively constant throughout the measurement. The average exciton energy saturates to a value that is 12 meV lower than the initial value within the first 20 ns. Based on these observations and using Eq.~\ref{eq:eqm}, we estimate inhomogenous and homogenous linewidths of CdSe QD from the ensemble measurements to be 14 and 25 meV, respectively, provided that Stokes shift of this sample has been measured to be 38 meV.~\cite{Akselrod2014} 

\begin{figure*}
\includegraphics{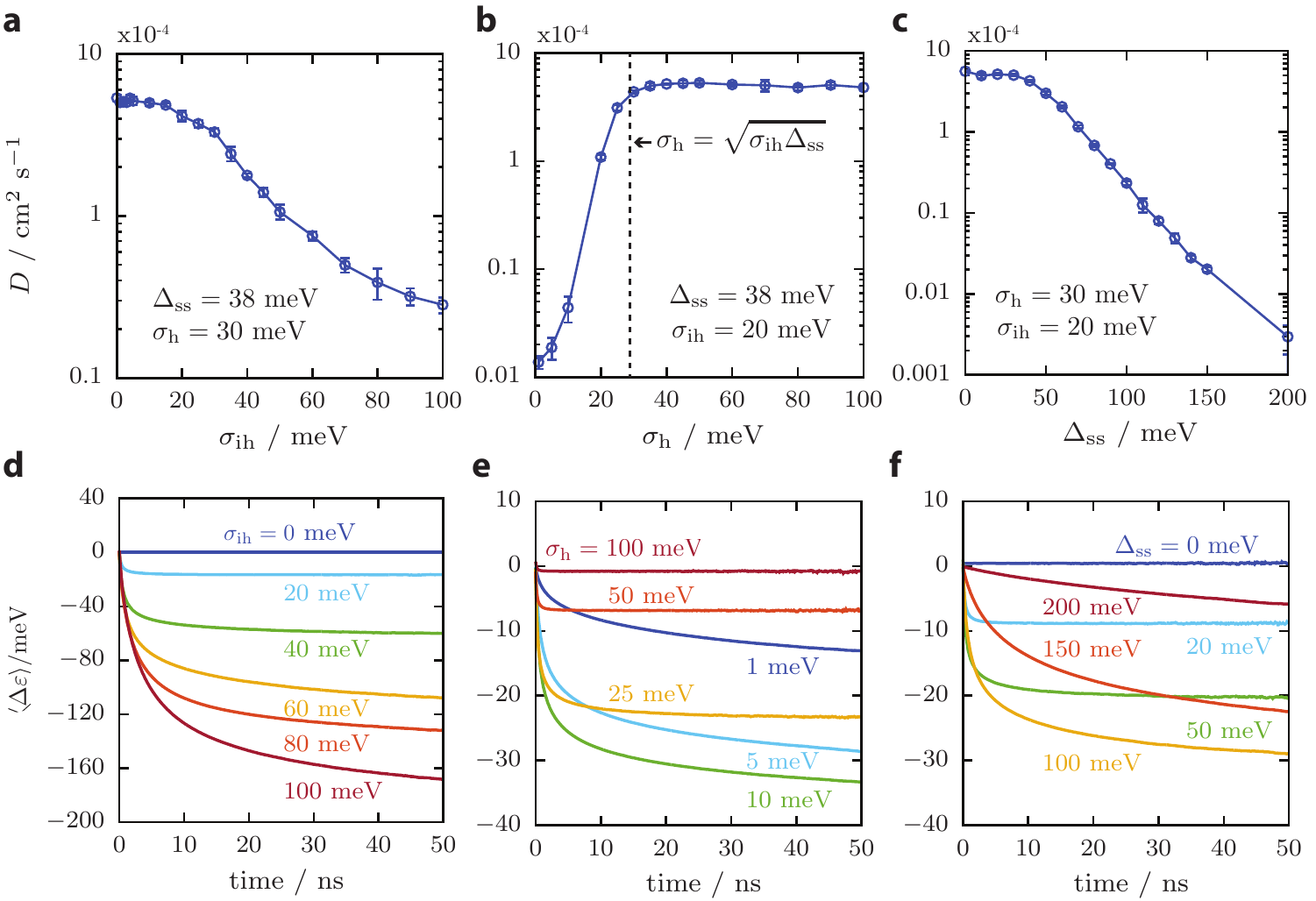}
\caption{Kinetic Monte Carlo simulations of the exciton diffusion in QD solid using F\"orster theory. (a,b,c) Exciton diffusivity versus inhomogeneous linewidth (a), homogeneous linewidth (b), and Stokes shift (c). (d,e,f) Dynamic redshift versus inhomogeneous linewidth (d), homogeneous linewidth (e), and Stokes shift (f). {Dotted line in (b) indicates the transition point beyond which mean exciton diffusivity remains constant.} For each parameter sweeps, other two variables that are held constant are indicated in the top row. In all cases, $\bar \varepsilon$ = 2.1 eV, $n$ = 1.7, $\eta$ = 1, and $\tau$ = 10 ns.}
\label{fig:fig_KMC}
\end{figure*}

The dynamic redshift of emission energy due to energetic disorder reveals information about the spatiotemporal dynamics of excitons in colloidal QD solids. Solving Eq.~\ref{eq:me} using a kinetic Monte Carlo algorithm,~\cite{Lee2015a} we relate the transient energetic to exciton diffusivity by varying inhomogeneous linewidth, homogenous linewidth, and Stokes shift as shown in Figure~\ref{fig:fig_KMC}. Since energetic disorder leads to a time-dependent diffusivity,~\cite{Lee2015a} we report mean exciton diffusivity to be the value determined when exciton population equilibrates to a thermalized distribution. Based on results plotted in Figure~\ref{fig:fig_KMC}, mean exciton diffusivity can be enhanced by decreasing the net loss in initial excitation energy, which can be achieved by reducing energetic disorder (inhomogeneous linewidth), increasing available thermal energy for exciton hopping (homogenous linewidth), and minimizing Stokes shift. {Figure~\ref{fig:fig_KMC}b highlights that increasing homogenous emission linewidth can mitigate the net negative effect of energy dissipation on mean exciton diffusivity until $\sigma_\mathrm{h} \approx \sqrt{\sigma_\mathrm{ih}\Delta_\mathrm{ss}}$, beyond which diffusivity remains constant.} For transient redshifts (Figures \ref{fig:fig_KMC}d-f), the final average energy follows the prediction by Eq.~\ref{eq:eqm} except in cases where $\sigma_\mathrm{h} \ll \sigma_\mathrm{ih}$ (Figure~\ref{fig:fig_KMC}e) and $\Delta_\mathrm{ss} \gg \sqrt{\sigma_\mathrm{h}^2 + \sigma_\mathrm{ih}^2}$ (Figure~\ref{fig:fig_KMC}f). In these situations, excitons never reach the thermal equilibrium because the probability of exciton hopping to neighboring QDs is lower compared to that of exciton decaying back to the electronic ground state. 

\section{Supertransfer for delocalized excitons}

 \begin{figure}
\includegraphics{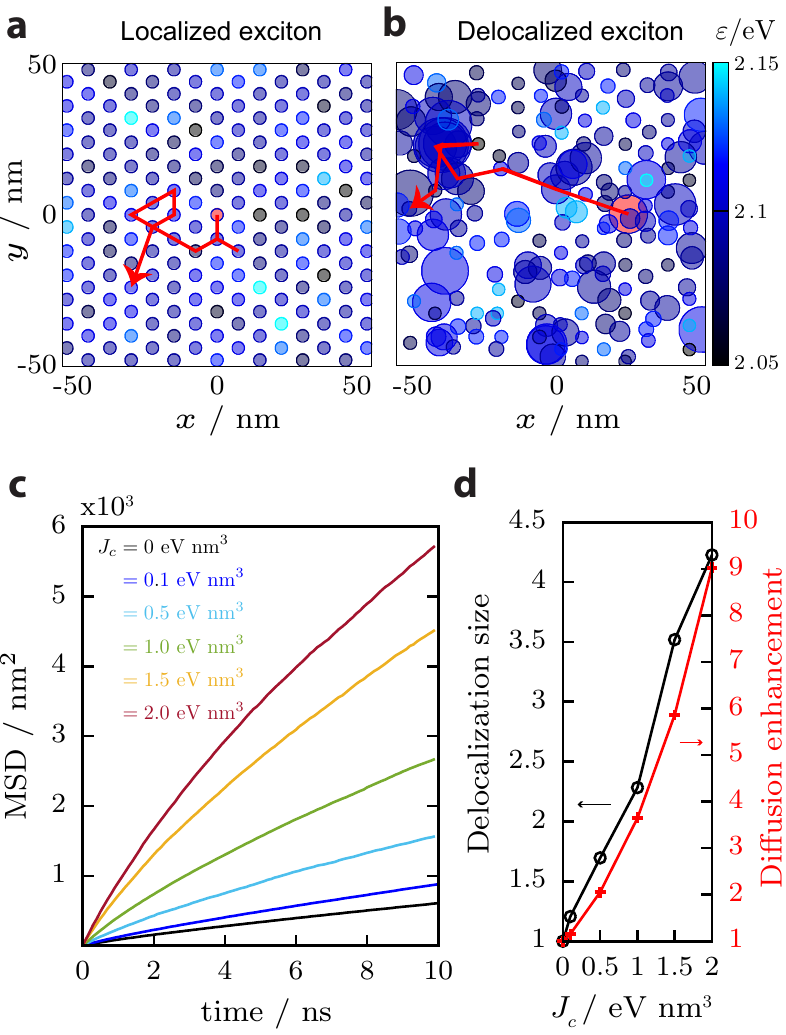}
\caption{Kinetic Monte Carlo simulations of exciton diffusion in electronically coupled QDs. (a,b) Exciton trajectories with $J_c=0$ (a) and $J_c = 0.1$ eV nm$^3$ (b). Each circle represents a delocalized exciton state whose color denotes its energy while the size of the circle is proportional to its inverse participation ratio (IPR). The initial position/state of the exciton is depicted by a red circle. (c) Mean squared displacements (MSD) of exciton with varying transition dipole coupling strength. (d) Exciton delocalization size and diffusion enhancement versus the transition dipole coupling strength. Exciton delocalization size is defined as the ensembled-averaged IPR, $\langle \text{IPR} \rangle$. The diffusion enhancement is defined as the ratio of delocalized exciton diffusivity to that of localized hopping. In all these studies, the system includes 2500 individual QDs, and we use $J_0 \equiv  \frac{\mu_0^2 }{4 \pi \epsilon \epsilon_0} = 0.1$ eV nm$^3$, $\varepsilon = 2.1$ eV, $\sigma_\mathrm{ih} = 20$ meV, $\sigma_\mathrm{h} = 30$ meV, $\Delta_\mathrm{ss} = 40$ meV, and $\tau = 10$ ns. }
\label{fig:fig_de}
\end{figure}

Due to inhomogeneous emission, low oscillator strength, and fast exciton dephasing,~\cite{Salvador2006,Leistikow2009,Accanto2012} excitons in II-VI QDs are thought to be localized on individual colloidal QDs with diffusion length measured between 5 to 35 nm.~\cite{Akselrod2014,Lee2015,Kholmicheva2015} Recent developments in colloidal QDs synthesis, however, have achieved ensemble emission linewidth as narrow as the homogenous linewidth, leading to highly ordered superlattices of colloidal QDs.~\cite{Boles2016,Weidman2018} By inducing favorable alignment of transition dipole moments of neighboring QDs, superlattice structure could offer unique optoelectronic properties of delocalized exciton. {Recent studies have also focused on enabling exciton delocalization by using electronically conductive surface ligands.~\cite{Crisp2013,Cohen2017,Azzaro2018}} If an exciton is delocalized over several QD sites, colloidal QD solid can achieve superradiance or superfluorescence as observed in molecular aggregates~\cite{Arias2013} and epitaxially grown QDs~\cite{Scheibner2007}. Most recently, superfluorescence has been reported in colloidal QD solids made from cesium lead halide perovskite (CsPbX$_3$, X = Cl, Br),~\cite{Raino2018} leading to speculations of enhanced exciton diffusion lengths in these systems through supertransfer.~\cite{Lloyd2010,Abasto2012,Geiregat2013} 

In previous sections, we have considered localized excitons, whose dynamics evolves via incoherent, hopping-type transport. In the presence of strong inter-QD interactions, however, electronic excitations can be delocalized across multiple QDs, leading to excitonic states that are superpositions of individual QD wave functions.  
Here we discuss the potential implication of exciton delocalization on the overall exciton transport. 

Let us consider the same QD model as elaborated in Section II. This time, we define the system Hamiltonian of $N$-number of quantum dots as,
\begin{equation}
H_s = \sum_n \varepsilon_n | n \rangle \langle n | + \sum_n\sum_{m > n}J_{n,m}  \left( | n \rangle \langle m | +  |m \rangle \langle n | \right),
\end{equation}
where $\varepsilon_n$ is the excitation energy of $n$-th QD, and $J_{n,m}$ is the transition dipole-dipole coupling given by,
\begin{equation}
J_{n,m} =J_c\frac{\kappa_{nm}}{d_{nm}^3},
\end{equation}
where $J_c$ is a coupling constant that scales the magnitude of the electronic coupling between neighboring QD. 
In the presence of environment-induced dephasing and energetic disorder, exciton transport involves two characteristic timescales: the short-time ballistic transport (\textit{i.e.}, MSD $\propto t^2$) and the long-time diffusive transport  (\textit{i.e.}, MSD $\propto t$).
Over a timescale longer than the exciton coherence time, the energy transfer process can be modeled as a series of hopping events among the eigenstates of the disordered system that diagonalize the system Hamiltonian,~\cite{Moix2013} 
\begin{equation}
H_s | \psi_i \rangle = E_i | \psi_i \rangle,
\end{equation}
where $\psi_i$ and $E_i$ are eigenvectors and eigenvalues of exciton state $i$. When $J_c=0$, we recover the localized exciton picture discussed previously. For $J_c > 0$, eigenstates of $H_s$ become delocalized over more than one quantum dot.

We model the change in spatiotemporal dynamics of excitons delocalized over multiple QDs using a kinetic Monte Carlo algorithm. Since the system is disordered and the diffusion constant depends only on the long-time dynamics, we adopt an analogous model to the multichoromophoric FRET.~\cite{Jang2004}  
In our model, exciton hopping from the donor (D) to the acceptor (A) eigenstate is captured by the generalized F\"orster theory described by the Fermi's golden rule,~\cite{Nitzan2006}
\begin{equation}
k_{\mathrm{D} \to \mathrm{A}} = \frac{2\pi}{\hbar}|V_{\mathrm{DA}}|^2 \Theta,
\end{equation}
where $V_{\mathrm{DA}}$ is the transition dipole-dipole coupling between the donor and the acceptor states, and $\Theta$ is the spectral overlap integral between normalized donor emission and acceptor absorption spectra. 
We calculate the electronic coupling via a line-dipole approximation,
\begin{equation}
V_{\mathrm{DA}} =  \frac{\mu_0^2 }{4 \pi \epsilon \epsilon_0} \sum_i^N \sum_{j \neq i}^N \frac{\kappa_{ij}}{R_{ij}^3} c_i^{(\mathrm{A})} c_j^{(\mathrm{D})},
\label{eq:Vda}
\end{equation}
where $c_i^{(\mathrm{A})} = \langle i | \psi_{\mathrm{A}} \rangle$ and $c_j^{(\mathrm{d})} = \langle j | \psi_{\mathrm{D}} \rangle$. 
By assuming that lineshapes of acceptor and donor eigenstates are Gaussians and that the magnitude of Stokes shift is the same among all eigenstates, we simplify the overlap integral expression as,
\begin{equation}
\Theta = \frac{1}{2\sigma_\mathrm{h}\sqrt{\pi}}\exp{\left[ \frac{-(E_\mathrm{A}+E_\mathrm{D} - \Delta_\mathrm{ss})^2 }{4 \sigma_\mathrm{h}^2} \right]}. 
\end{equation}

We quantify the extent of delocalization of $n$-th eigenstate by using the inverse participation ratio,
\begin{equation}
\text{IPR} = \frac{1}{\sum_i^N \langle i | \psi_n \rangle}.
\end{equation}
For instance, if the eigenstate is symmetrically delocalized over two QDs $i$ and $j$, \textit{i.e.}, $|\psi_n \rangle = 1/\sqrt{2}(|i\rangle + |j\rangle)$, then IPR=2. 

Figures \ref{fig:fig_de}a and \ref{fig:fig_de}b depict kinetic Monte Carlo trajectories of localized and delocalized excitons, respectively. In the case of localized hopping, exciton hopping events to nearest neighbor sites are most common. However, in the presence of strong inter-QD electronic coupling, an exciton may hop to a state that is energetically resonant but spatially far away due to the enhancement in the net transition dipole moment.
Therefore by increasing $J_c$, we find that exciton delocalization enhances exciton diffusion (Figure~\ref{fig:fig_de}c). Based on Figure \ref{fig:fig_de}d, we confirm that diffusivity increases as the average IPR of excitonic states formed in the coupled QD solid increases. Ideally, if all QDs have parallel transition dipole moments, and if both donor and the acceptor states are symmetrically delocalized across $M$-number of QDs, then,
\begin{equation}
 k_{\mathrm{D} \to \mathrm{A}}\text{(delocalized)} = M k_{\mathrm{D} \to \mathrm{A}}\text{(localized)}.
\end{equation}
In reality, however, due to random orientation of the transition dipole moments of individual QDs as well as the inhomogeneous broadening of QDs, the overall diffusion enhancement factor, defined as $D_{\mathrm{delocal}}/D_{\mathrm{local}}$, is less than the theoretical maximum of $D_{\mathrm{delocal}}/D_{\mathrm{local}}=\langle \mathrm{IPR} \rangle^2$. 

We have found that an order-of-magnitude increase in exciton diffusivity is expected even when the average exction delocalization size is only $\sim 4$ QDs. Therefore, as long as there is sufficiently strong electronic coupling between a few neighboring QDs, there is a possibility of significantly enhanced exciton transport---even without perfect alignment of transition dipole vectors. 

\section{Conclusions and Outlook}
Disorder is an intrinsic property of QD solids and can manifest as spatial variations in the excitation energy, otherwise known as inhomogeneous broadening. 
In general, excitation of inhomogeneously-broadened ensemble of QDs leads to an energetic relaxation of exciton population, which results in a mean exciton diffusivity that is lower than expectations for a perfectly ordered material. Our work highlights two directions in which mean exciton diffusivity in disordered QD solids can be increased. 
First, exciton diffusivity can be maximized by specifically balancing the effects of homogeneous broadening, inhomogeneous broadening, and Stokes shift.
Second, under ambient conditions, an order-of-magnitude increase in exciton diffusivity can be achieved by harnessing the effects of exciton delocalization. Notably, we have observed that the remarkable energy transport enhancements arising due to exction delocalization are robust against disorder in the energetic and orientational arrangements of QDs in the material. This observation motivates continued efforts to achieve exciton delocalization by controlling QD-QD coupling strength either at the individual QD level through surface chemistry or at the collective level by fabricating highly ordered QD arrays.

\begin{acknowledgments}
E.M.Y.L acknowledges helpful discussions with Hendrik Utzat. The submission of this work has been supported by the 2017 AIChE Annual Meeting's Electronic and Photonic Material Graduate Student Award, sponsored by Journal of Vacuum Science and Technology.
This work has been funded by the Center for Excitonics, an
Energy Frontier Research Center funded by the US Department
of Energy, Office of Science, Office of Basic Energy
Sciences, under Award DE-SC0001088 (MIT)
\end{acknowledgments}

\bibliography{ref}

\end{document}